# The Physics of Gudeg: Learning the Mechanics and Thermal Properties Using Collaborative Project based Activities for the High School Physics

**Bayu Setiaji** [1,*], **Pramudya Wahyu Pradana**[1], **Febrina Siska Widyaningtyas**[1], **Purwoko Haryadi Santoso**[2], **Yusman Wiyatmo**[1], **Heru Kuswanto**[1]

[1] Department of Physics Education, Universitas Negeri Yogyakarta, Sleman 55281, Indonesia

[2] Department of Physics Education, Universitas Sulawesi Barat, Majene 91413, Indonesia

**\* Correspondence:** bayu.setiaji@uny.ac.id

**Abstract**

In this paper, we describe a presentation on the physics of Gudeg, a traditional food from Indonesia specifically originated in the Special District of Yogyakarta. This learning context is designed for the high school physics curricula. The physics presentation focuses on the making processes of Gudeg. Qualitative interviews with Gudeg makers were carried out by the researchers to thematize the process of making Gudeg and highlight its educational connections for the physics learning. Five extracted learning themes are how the density concept behind peeling the jackfruit skin (main Gudeg ingredient), how the relation between the Young's modulus concept and the jackfruit sections, how the texture-torque experiment of the sweet and tasty gudeg, how the effect of boiling mechanism on the texture of the jackfruit, and how the conduction and convection of the preserved Gudeg. Using our learning strategy which is so-called "Collaborative Project based Teaching", we provide simple experimentations and demonstrations of the physical concepts behind these Gudeg processes that are promising for conceptual physics learning by managing triple educational roles between teachers, students, and Gudeg practitioners. This approach can be generally adopted beyond physics to promote the excitement of traditional knowledge which can enhance pedagogical approach in educational setting.

**Introduction**

Indonesia is well known as a nation with the diversity of cultural heritage. The indigenous knowledge, system, and custom is our resources that hold significant value for the Indonesian people. This also reflects the ethnology inherent in our society and cultural communities (Deta et al., 2024). Our national artefacts manifest in various states including but not limited to values, houses, songs, dances, attires, historical sites, games, and foods. These enormous cultural entities are subsequently deployed as our local wisdom. We believe that presenting local knowledge for physics learning could also be adopted beyond the Indonesian context since every nation should have wonderful values. Hence, our closest circumstances should be the best avenue to approach physics education.

The novel idea of incorporating indigenous knowledge to enhance education has garnered increasing attention including in the physics education research (Deta et al., 2024; Khery et al., 2021; Martawijaya et al., 2023; Sari et al., 2020; Susanto et al., 2023). The interplay between physics and culinary contexts has been explored through former studies. These studies highlight how culinary activities can effectively convey scientific principles, making science more accessible and engaging (Fuller et al., 2022; Mathijssen et al., 2023; Rowat et al., 2014; Vilgis, 2015). This delivery model offers student to unique learning experiences and supports their conceptual understanding of physics (Deta et al., 2024; Khery et al., 2021; Surtikanti et al., 2017; Susanto et al., 2021, 2023). Cultural artefacts, as mentioned above, are potentially engaged in the learning process (Deta et al., 2024; Khery et al., 2021; Sari et al., 2020; Susanto et al., 2023). To the best of our knowledge, there is a notable scarcity of research exploring the context of traditional foods. Therefore, further research is a necessity since the culinary heritage also holds potential role to support equality and the culturally relevant pedagogy.

In this paper, traditional food is conceptualized as a culturally relevant content and as a meaningful context for constructing scientific understanding. The instructional design draws upon cognitive development theories, particularly Piaget's constructivist view and Vygotsky's socio-cultural perspective. Piaget posits that learners develop cognitive structures through interaction with their environment, and traditional food practices function as socio-cultural activities where such interactions naturally occur (Setiaji & Jumadi, 2018). Engaging with these practices in the classroom allows students to build physics understanding through lived experience (Dewi & Primayana, 2019). Vygotsky, meanwhile, emphasizes that knowledge is socially mediated, constructed through collaboration with more knowledgeable others, including parents, teachers, or peers. In the context of this paper, community members can act as valuable learning agents. Wertsch (1993) highlights the role of both physical and psychological tools in learning, suggesting that indigenous knowledge, like traditional foods, can serve as instructional tools alongside conventional instruments. Other scholars argue that physics education can link students' cultural experiences to formal physics concepts as supported by the curriculum, thereby grounding scientific learning in familiar and relevant contexts (Mandikonza, 2019).

The focus of this paper is to explore one of Indonesia's most storied local delicacies, namely Gudeg (Figure 1). As a cornerstone of traditional Indonesian cuisine, Gudeg originated in the Special Region of Yogyakarta and has remained a dietary staple since the era of the Mataram Sultanate circa 1755 (Yudhistira, 2022). The dish is primarily composed of young jackfruit ("nangka muda" in Indonesian language) simmered in coconut milk and seasoned with a precise blend of palm sugar, salt, herbs, and spices. The result is a distinctively sweet and savory profile that has solidified Gudeg's status as a native culinary icon of the Yogyakarta community. Beyond the jackfruit itself, a complete serving of Gudeg is traditionally accompanied by an array of side dishes, including "krecek" (spicy cow skin stew), hard-boiled eggs, and chicken. This combination reflects the Javanese philosophy of harmony, balancing the sweetness of the fruit with savory and spicy accents. Today, the Gudeg transcends its historical roots as royal fare, serving as both a daily meal for locals and a must-try cultural experience for international travelers.

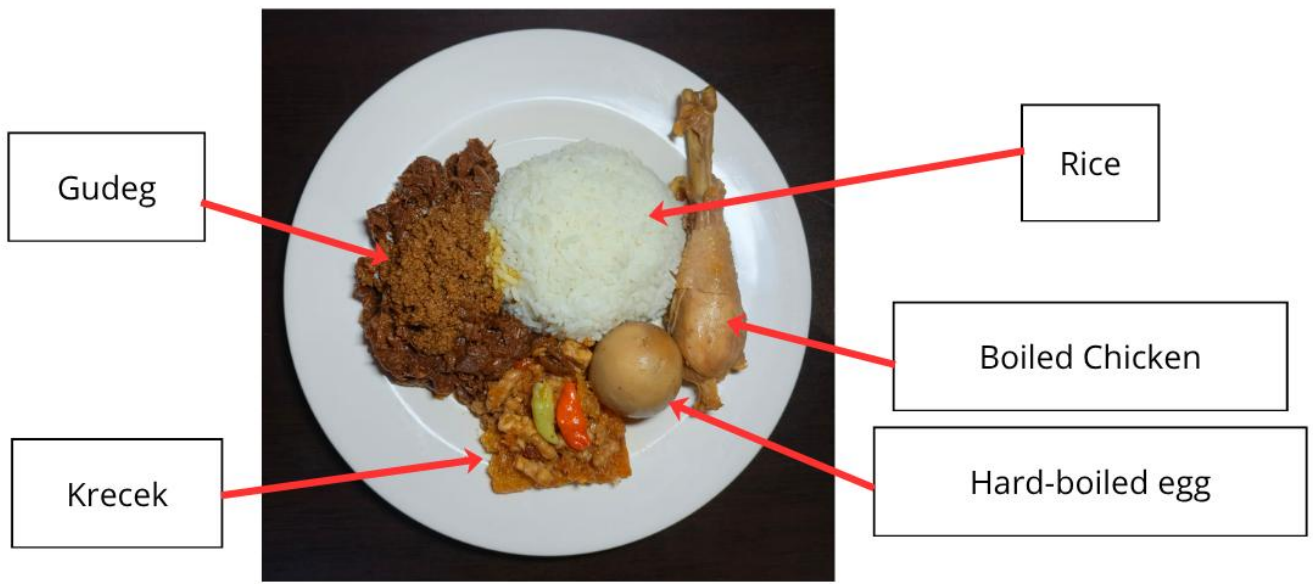

**Figure 1**. A complete serving of Gudeg.

Figure 2 presents the overall process of Gudeg production (Yudhistira, 2022), which serves as the contextual pathway for discussing its related physics concepts in this paper. The process begins with peeling and chopping the young jackfruit, followed by mixing with water, coconut milk, and spices, and continues through the boiling stage. The final stage involves preservation, in which Gudeg is packaged in two forms, dry and wet versions. For dry Gudeg, preservation involves a canning process that includes filling cans with Gudeg and complementary ingredients, followed by an exhausting step at approximately 90°C to remove entrapped air from the container and food matrix. Heating during exhausting stage causes gas expansion and release, and subsequent cooling leads to gas contraction that forms a partial vacuum inside the can. The sealed cans are then sterilized at high temperature and pressure (typically 121°C at around 2 bar) and finally cooled in a water tank to stabilize the temperature. In contrast, wet Gudeg does not undergo this preservation type, as it is typically prepared for immediate or shorter-term consumption and distributed directly without vacuum packaging.

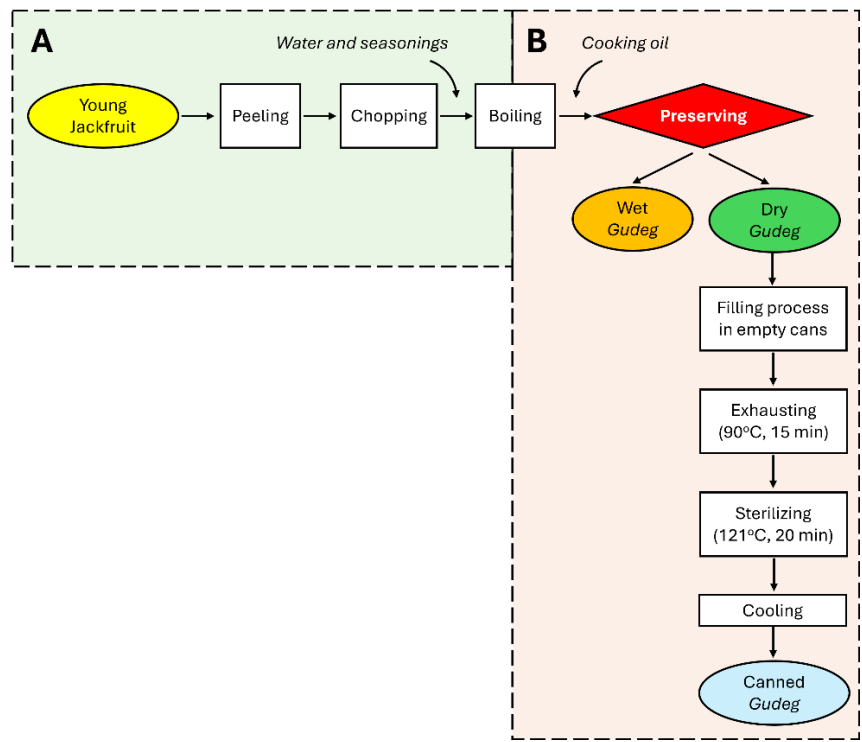

**Figure 2**. The systematic process of producing a Gudeg. (A) Focused on the cooking process of Gudeg from the primary ingredient, young jackfruit. ( B) Focused on the preserving the Gudeg which can be consumed for a prolonged time.

This paper is intended to explain physics learning we designed behind the Gudeg phenomena which is geared toward nonspecialist and the general audience including high school physics students. Our learning strategy is designed to support collaborative project-based environment. Table 1 summarizes the steps of Gudeg making processes, collaborative project-based activities, and main physical concepts that can be studied through the Gudeg knowledge. Each step would be represented using project-based collaborative experiments so that students are interactively engaged with physics in the classroom. The model is also expected to enhance their senses of learning agency and self-belonging that are critical for effective learning.

Table 1 is our basis for structuring this paper. This matrix is based on our former study using qualitative interviews and direct observations with two main Gudeg producers as our primary case studies. They are "Bu Rini" and "Bu Tjitro" who are chosen to represent two different approaches of the Gudeg preservation. The qualitative data was then analyzed using a thematic analysis approach to find the connection between the Gudeg context and the physical phenomena. Our findings discover that Gudeg is often packaged for retailers in dry and wet states. The "Bu Rini" factory represents the dry Gudeg and the "Bu Tjitro" factory produces the wet version of Gudeg. They are invited to this research because we argue that education should have cross-sector connection when our teaching strategy is collaborative project-based learning. We build this argument using socio-cultural theory as formerly reported by Vygotsky's works. Adapting socio-cultural theory, education could be considered as a social community. When we view education as a learning organization through this theory, the school is not an isolated factory producing credentials and should be a growing hub within a larger socio-economic network. In this model, the real sectors as demonstrated by Gudeg producers are no longer just external observers, they are definitive stakeholders whose survival and innovation are directly linked to educational output. In this paper, students' project will be tailored collaboratively with the Gudeg practitioners for the collaborative project-based learning model. By engaging directly with the practitioners, the learning projects bridge the gap between theoretical classroom concepts and the lived realities of Yogyakarta's traditional food industry. Ultimately, this collaborative model aims to foster a deeper sense of cultural ownership among students while documenting the evolving role of Gudeg in a modern landscape of physics education.

**Table 1**. Main processes of cooking Gudeg , collaborative project-based activities, and related physical concepts

| Cooking steps | Collaborative project-based activities | Main physical concepts |
|---|---|---|
| Peeling | The Density Analysis of Different Jackfruit Sections | Mass density and Archimedes principle |
| Chopping | The Elasticity Experiment between Jackfruit's Skin, Cortex, Perianth, and Core Sections | Elasticity and Young's modulus |
| Mixing | The Texture-Torque Experiment of the Sweet and Tasty Gudeg | Torque Viscocity |

| Cooking steps | Collaborative project-based activities | Main physical concepts |
|---|---|---|
| Boiling | The Effect of Boiling Mechanism on The Texture of the Jackfruit | Thermal effect |
| Preserving | The Convection of Wet Gudeg and Conduction of Dry Gudeg | Heat transfer (conduction, convection) |

**Engaging Students' Activity for Collaborative Project-Based Environment**

Collaborative project-based teaching links the real context of physical phenomena with the students and their learning. This model integrates collaborative aspects and students' activities during the project construction and experiments. In summary, the illustration of our proposed learning strategy can be depicted in Figure 3 below. The socio-cultural framework became the core theoretical standpoint to design this model to use the Gudeg making process (Figure 2) as our learning path to study physics. The basis of our developed learning model was defined by former work (Pujianto et al., 2024). The model is structured into five instructional phases that facilitate contextual learning by connecting students' cultural experiences with core physics concepts. Each phase is designed to encourage active exploration (Deta et al., 2024; Khery et al., 2021).

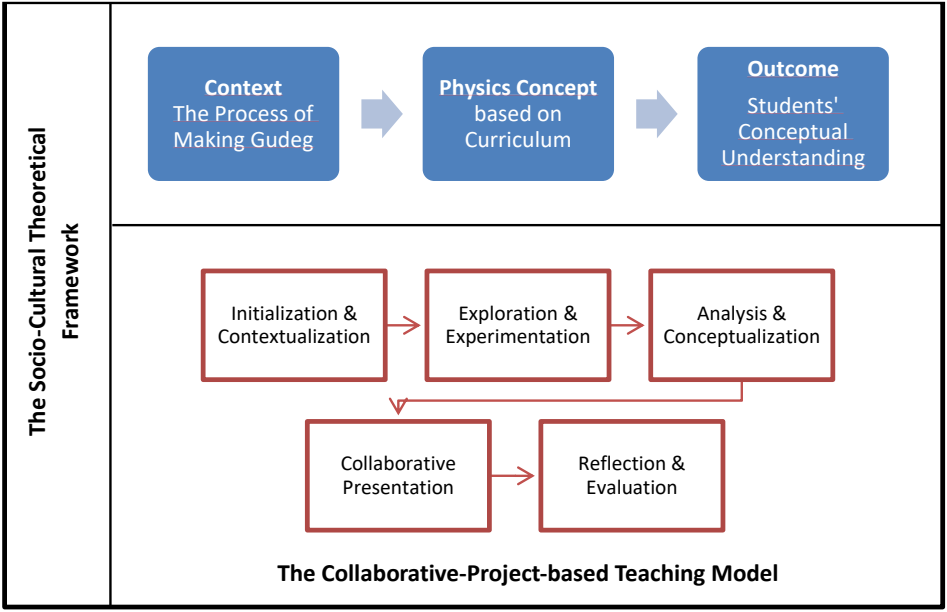

**Figure 3**. The Collaborative-Project-based Teaching Model.

The first step, initialization and contextualization introduce traditional food as a learning context and identifying relevant physics concepts through initial discussions. The learning then continues with exploration and experimentation to observe real-life food processing and conduct simple classroom experiments related to observed physics concepts. After the experiments, students analyze their data and link it to physics concepts through guided reflection and mathematical modeling. Teachers then guide collaborative students' groups to present the results of their investigations in a collaborative and communicative setting. Eventually, the classroom end the activity with reflection and evaluation to review the learning process and teachers can assess students' understanding of physics using cultural approach.

To manage these learning activities throughout the collaboration within the classroom, we can use a triple learning roles between teachers, students, and Gudeg producers. Teacher role acts as a technical director ensuring the educational objectives are met and the project remains structured. Student role acts as a consultant learner analyzing observations and interpreting the findings into practical advice. Practitioner role acts as the subject matter expert validating if the students' findings match the traditional standards and cultural essence of a well-cooked Gudeg. This collaborative framework ensures that academic curiosity is balanced with real-world experience, creating a space where modern learning respects and preserves traditional wisdom. By working together, all participants contribute to a deeper understanding of how local heritage can be documented and sustained for future generations.

**Experiment 1: The Density Concept behind Peeling the Jackfruit Skin**

Gudeg is made from young jackfruit (*Artocarpus heterophyllus*), locally known as "nangka muda" or "gori" (Yudhistira, 2022). After peeling, the jackfruit is divided into three main sections (Figure 4): cortex, perianth, and core (González-Regalado et al., 2024). From a physics perspective, each section can be treated as a composite material with specific mass (density) and internal structure, which influences its behavior during cooking.

We have an inspiring idea for physics experiments to engage students' projects while studying the young jackfruit sections. To integrate socio-cultural theory into this experiment, we treat the jackfruit sections as biological engineering objects. The Gudeg producers provide the expertise on which sections absorb flavors best (anecdotal evidence), while the students provide the empirical data to explain why through the concepts of mass density and structural anisotropy.

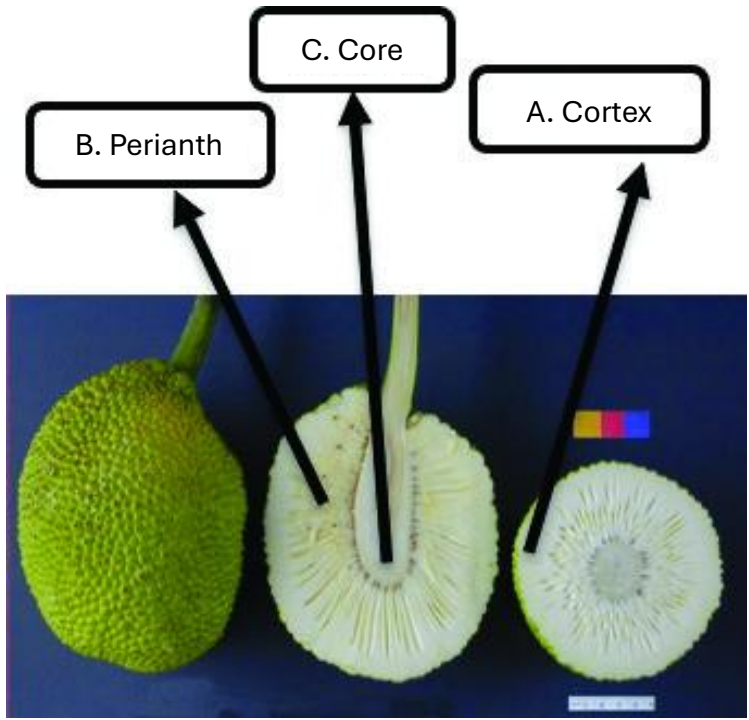

**Figure 4**. The Learning Sections of Main Gudeg Material, Young Jackfruit (*Artocarpus heterophyllus*). A. Cortex. B. Perianth. C. Core. (González-Regalado et al., 2024)

Students conduct a simple experiment to determine the density ($\rho$) of each jackfruit section using water displacement based on Archimedes' principle, as illustrated in Figure 5. By measuring mass ($m_1$) and volume ($V_1$ and $V_2$), students analyze density differences and relate them to buoyancy and porosity. These variations explain why the perianth absorbs flavors more rapidly, while the denser core requires longer thermal processing.

# Experiment 1

**1** Weigh Sample ($m_1$)

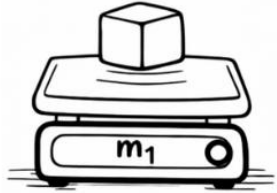

Measure the mass of each jackfruit section to quantify its material content.

**2** Measure Initial Volume ($V_1$)

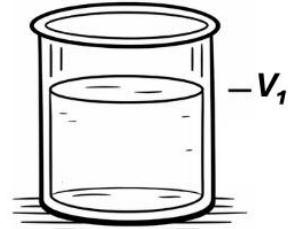

Record the initial water level before immersion as a reference volume.

**3** Immerse in Water

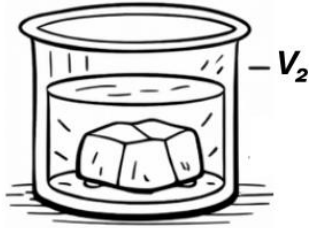

Submerge the sample to determine displaced volume ($\Delta V = V_2 - V_1$) based on Archimedes' principle.

**Calculate Density**

Density ($\rho$) is defined as mass per unit volume, indicating how compact or porous the material is.

**Figure 5**. Density Measurement of Jackfruit Sections Using Water Displacement

**Experiment 2: The Young's Modulus Concept and the Jackfruit Sections**

The anatomical structure of young jackfruit provides a useful context for studying structural mechanics. Because each anatomical section, i.e. the skin, cortex, perianth, and core, features distinct fiber orientations and cellular densities, the fruit behaves as an anisotropic biological material that responds differently to the thermal and mechanical stresses of Gudeg production. By treating these sections as biological beams, students can study the concept of Young's Modulus ($E$) while learning the Gudeg chopping process.

Young jackfruit can be chopped into different sections. Since each section of young jackfruit has different fiber orientations, we can use this mechanical property to explore Gudeg further to study physics. Jackfruit fibers are aligned differently in each section, affecting how the fruit shreds when we cook. In this experiment, we apply Young's Modulus concept. It is a measure of a mechanical stiffness of material to the botanical structures of young jackfruit. By assuming the fruit sections as biological beams, students can quantify how much force is required to cause a specific amount of deformation (strain) before the cellular walls of the jackfruit give way.

Young's Modulus ($E$) is defined by the ratio of tensile stress ($\sigma$) to tensile strain ($\epsilon$):

$$E = \frac{\sigma}{\epsilon} = \frac{F/A}{\Delta l/l_0}$$

$F$ : The force applied (N).

$A$ : Cross-sectional area of the jackfruit sample (m²).

$\Delta l$ : The change in length (stretching/ compression) (m).

$l_0$ : The initial length before stretching/ compressing (m).

Students investigate the mechanical properties of jackfruit sections by applying tensile force and measuring deformation, as illustrated in Figure 6. By measuring the applied force (F), cross-sectional area (A), and change in length (ΔL), students determine stress and strain to evaluate Young's modulus (E). This allows comparison of elasticity among different sections.

In this learning context, the elastic phase of young jackfruit is the point where the jackfruit can be squeezed or pulled and still return to its original shape. Once it enters the plastic phase, the fibers have permanently shifted which is exactly what happens during the long boiling process of making Gudeg.

# Experiment 2

**1) Clamp Jackfruit**
Secure the jackfruit sample to maintain a fixed initial length ($L_0$) for controlled mechanical testing.

**2) Apply Force (F)**
Apply a tensile load to the sample to generate stress based on its cross-sectional area.

**3) Measure Deformation (ΔL)**
Measure the change in length to quantify strain and evaluate the material's elastic response.

## Young's Modulus Concept

Young's Modulus ($E$) describes material stiffness as the ratio of stress to strain, indicating resistance to deformation.

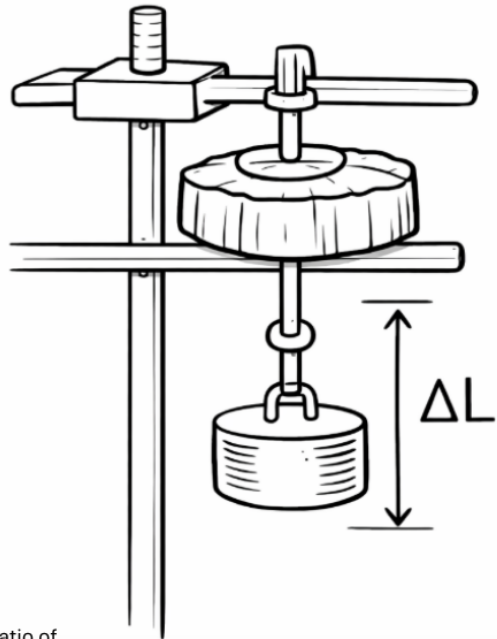

**Figure 6**. Measurement of Young's Modulus of Jackfruit Sections

**Experiment 3: The Texture-Torque Experiment of the Sweet and Tasty Gudeg**

Mixing in Gudeg preparation involves the interaction between mechanical force and fluid resistance. Proper stirring is required to achieve a homogeneous mixture, while excessive force may damage the jackfruit structure. We can design a texture-torque correlation experiment. This experiment focuses on the viscoelasticity of the jackfruit and the fluid dynamics of the coconut milk reduction. The laboratory aims to determine the critical agitation point. This is a physical measure from the exact cutoff point when the mechanical force of stirring begins to damage the jackfruit's cellular structure instead of aiding flavor distribution.

In this experiment, students investigate the relationship between applied force and fluid resistance during stirring, as illustrated in Figure 7. By applying a force (F) at a certain radius (r), the resulting torque ($\tau = r \times F$) represents the mechanical effort required to mix the material. As heating progresses, the mixture becomes more viscous due to evaporation and concentration of components. This increase in viscosity (η) leads to greater resistance to flow, requiring higher torque to maintain stirring.

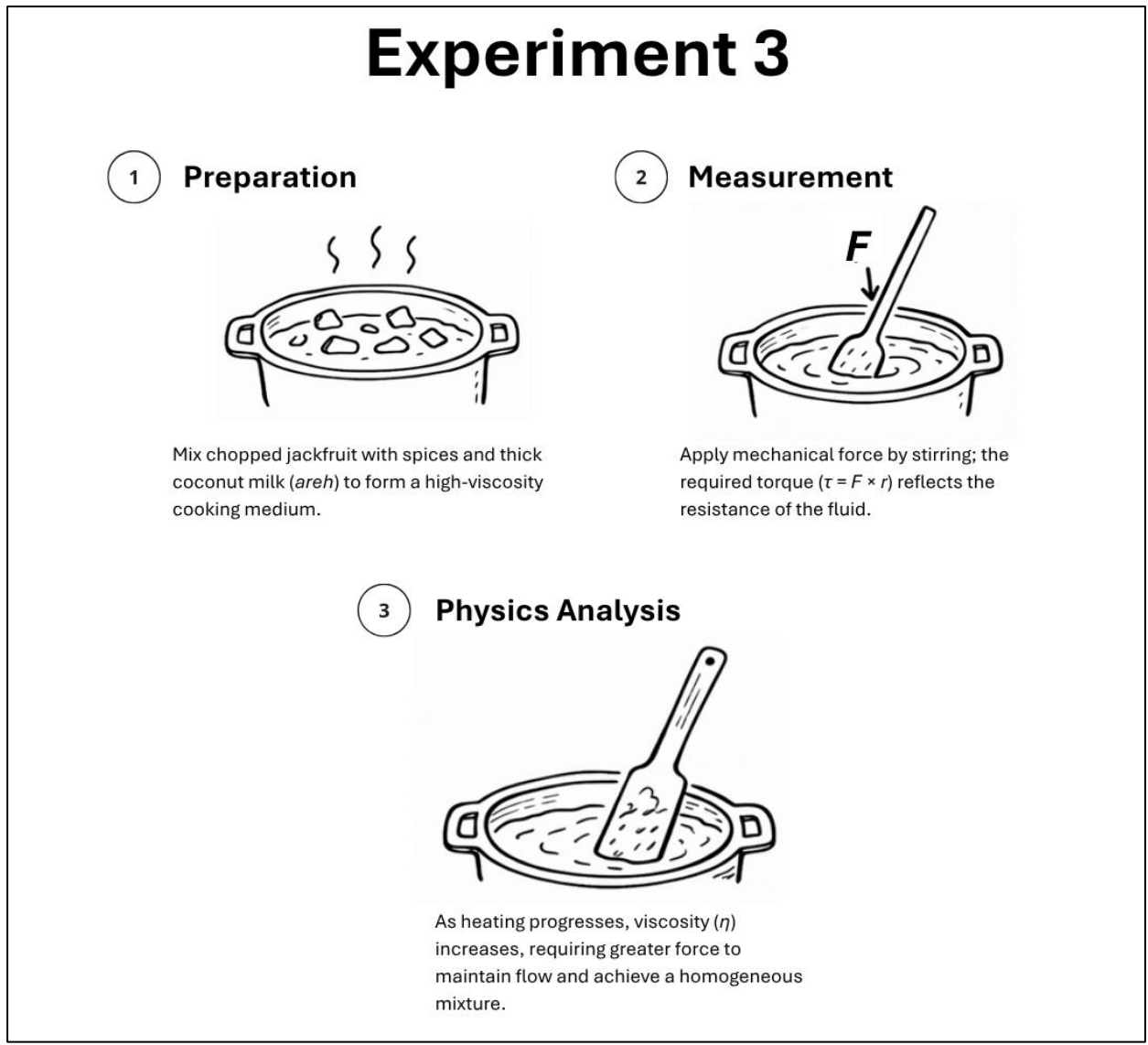

**Figure 7**. Relationship between Torque and Viscosity in Gudeg Mixing

At a certain stage, excessive mechanical force may exceed the structural strength of the softened jackfruit, indicating a critical agitation point where further stirring can damage the material.

**Experiment 4: The Effect of Boiling Mechanism on The Texture of the Jackfruit**

This experiment investigates how thermal treatment affects the mechanical properties of jackfruit. After boiling at a constant temperature, the structural integrity of each section is evaluated through mechanical response. The experiment goal is to see how much kinetic energy each section can absorb before its cellular structure, already weakened by heat, undergoes permanent deformation. This experiment bridges the gap between the feeling of a chef's fingers and the cold, hard data of physics.

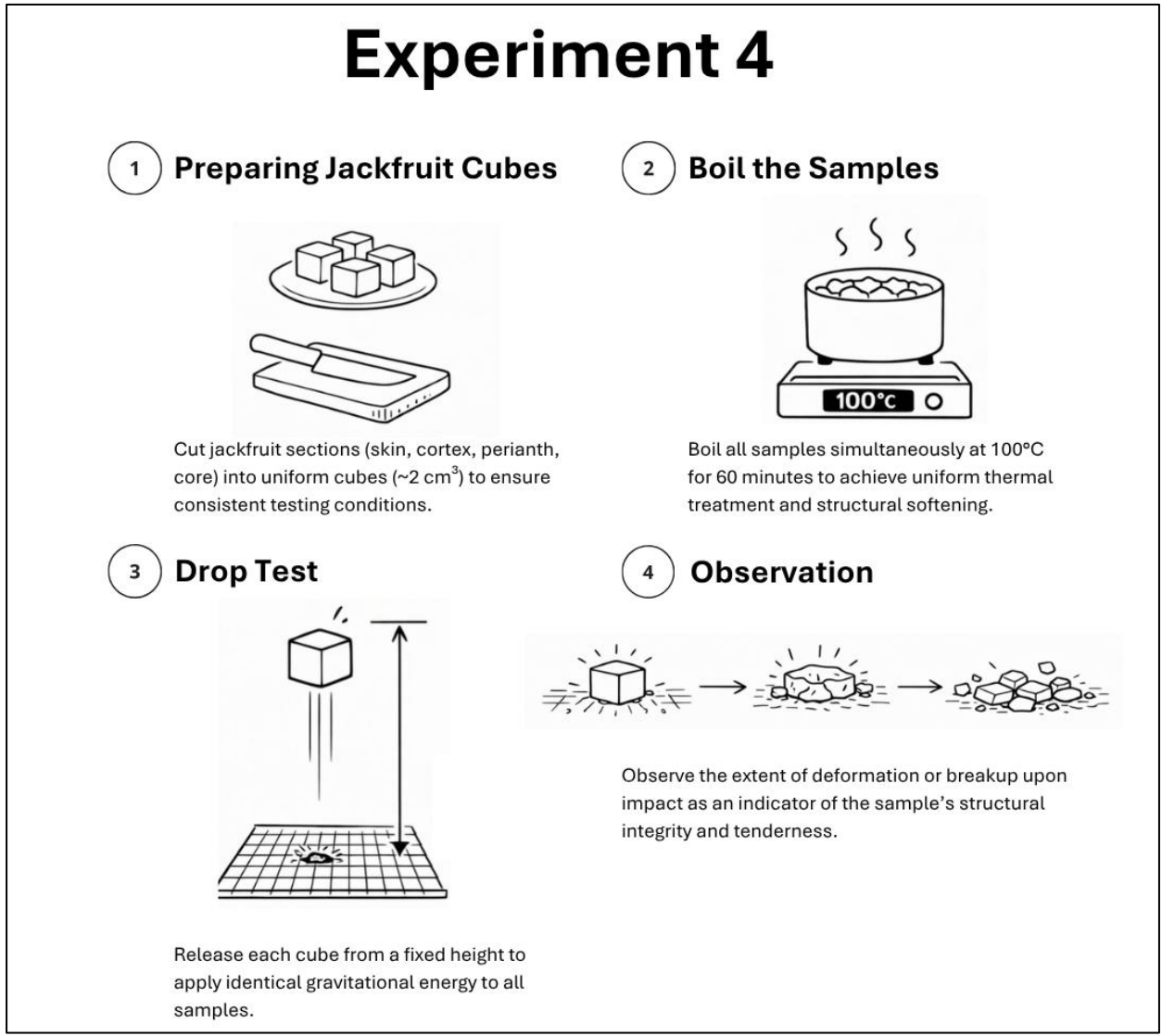

**Figure 8**. Effect of Thermal Treatment on Jackfruit Texture Using Drop Test

Students prepare uniform jackfruit cubes (~2 cm³) and boil all samples simultaneously at 100°C for a fixed duration to ensure consistent thermal conditions, as illustrated in Figure 8. After boiling, each sample is released from a fixed height onto a flat surface, providing the same gravitational potential energy ($E = mgh$) to all samples. The resulting deformation or fragmentation upon impact is used as an indicator of texture. For more detailed analysis, deformation is recorded using slow-motion video, and the vertical compression is quantified by measuring the change in sample height before and after impact.

The drop test reflects the material's ability to absorb energy before undergoing deformation. Softer samples exhibit greater plastic deformation, while stiffer samples retain their shape or rebound. Typically, the perianth shows greater deformation due to higher water absorption, whereas the denser core remains more structurally intact. This explains differences in tenderness and required cooking duration.

**Experiment 5: The Convection of Wet Gudeg and Conduction of Dry Gudeg**

This experiment compares heat transfer mechanisms in wet and dry Gudeg during thermal processing. Students prepare two systems representing wet Gudeg (low-viscosity broth) and dry Gudeg (high-viscosity paste), as illustrated in Figure 9. Both samples are heated simultaneously in a water bath at 100°C to ensure identical external conditions.

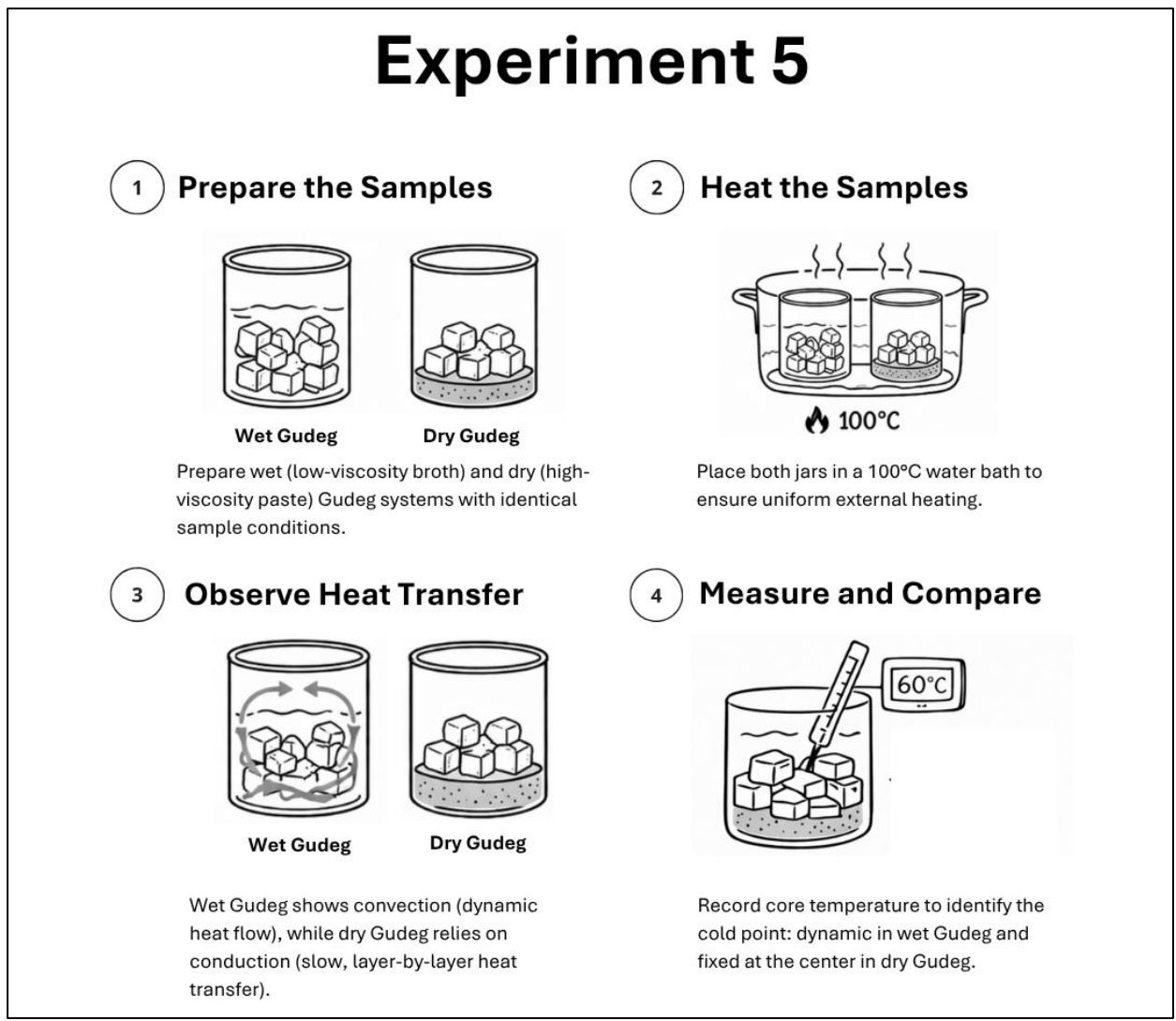

**Figure 9**. Heat Transfer Mechanisms in Wet and Dry Gudeg

In the wet system, a tracer (e.g., food coloring) is added to visualize fluid motion. As heating progresses, the fluid circulates due to density differences, forming convection currents that transport heat rapidly throughout the sample. In contrast, the high viscosity of dry Gudeg suppresses fluid motion, and heat transfer occurs primarily through conduction, where energy propagates slowly through the material without bulk movement.

By measuring the temperature at the core, students identify the cold point, defined as the location that takes the longest to reach the target temperature. This point is dynamic in wet systems due to fluid circulation but remains fixed at the center in dry systems. This difference explains why dry Gudeg requires longer heating to achieve uniform thermal penetration.

**Adaptation to the Indonesian Physics Education Curriculum**

The current Indonesian physics education curriculum promotes the vision of culturally relevant pedagogy for the more equitable and accessible learning. The big idea emphasize on opportunities for students to develop themselves according to their interests and talents (Fauzan et al., 2023). The curriculum is designed with greater flexibility, focusing on essential core content (Saputra et al., 2023). The main emphasis of this curriculum is on student character development (Fauzan et al., 2023). There are minimum competency standards for each subject, including physics, mandated by the curriculum to ensure that these standards must be met by all schools across Indonesia.

Table 2 summarizes the strategic intersection of our collaborative project-based teaching model with the specific learning outcomes mandated by the Indonesian national curriculum. In the Indonesian education system, Physics is a core scientific discipline studied by students in Grades 10 through 12. The current curriculum divides the secondary physics journey into two distinct developmental phases over three years to ensure smooth progression from general science to specialized inquiry. The first stage, Phase E, is implemented in Grade 10 and serves as a critical pedagogical bridge, transitioning students from the foundational science of junior high school to the more rigorous analytical demands of senior high school. During this phase, the curriculum avoids overly complex mathematical abstractions, focusing instead on introductory and basic topics such as measurement, unit consistency, and the fundamental laws of energy that underpin everyday phenomena. This allows students to build confidence by applying simple physics to the familiar context of Gudeg production, such as quantifying the dimensions of jackfruit anatomy using standardized tools.

As students' progress into Phase F, which spans Grades 11 and 12, the curriculum shifts toward a deeper, more specialized exploration of physical laws. In this phase, the content is intentionally streamlined to cover only essential or core materials such as fluid dynamics, thermodynamics, and the mechanics of materials rather than an exhaustive list of disparate topics. This less is more approach is specifically designed to provide students with the necessary time wealth to engage in intensive, project-based activities that require higher-order thinking skills. Within our model, this phase allows students to dive into five experimental designs. By focusing on these essential concepts, Phase F encourages students to move beyond classical approach where they must interpret raw data to solve the practical preservation challenges faced by Gudeg practitioners.

**Table 2**. Mapping Learning Outcomes of the Indonesian Physics Education Standards with the Collaborative-Project based Activities using Cultural Context

| Grade/ Phase | Learning outcomes based on the curriculum | Physics content | Learning activity |
| --- | --- | --- | --- |
| Grade 10 (Phase E) | Students can design and carry out scientific inquiries to solve environmental or local problems. | Measurement and analysis | Measuring dimensions and mass of core vs. perianth sections to determine the mass density ($\rho$). |
| Grade 10 (Phase E) | Students understand the concept of energy, heat, and their transformations in daily life. | Thermodynamics (heat transfer) | Observing convection and conduction using tracers in wet vs dry Gudeg to map heat flow. |
| Grade 11 (Phase F) | Students can analyze the characteristics of fluids and apply them to technological contexts. | Fluid dynamics & rotational dynamics | Measuring the angle of deflection (torque) of the stirring paddle as coconut milk reduces and viscosity increases. |
| Grade 11 (Phase F) | Students can analyze the relationship between stress, strain, and the elasticity of materials. | Elasticity (Young's modulus) | Performing compression tests on jackfruit to calculate Young's Modulus and identify the critical agitation point. |
| Grade 11 (Phase F) | Students apply principles of thermal energy and mechanics to analyze material properties and structural changes. | Thermodynamics | Investigating how heat and fluid absorption create a softening gradient between the dense core and the porous perianth sections. |

In a classroom of 36 students, managing five distinct experiments requires a station-based management style. Table 3 presents our proposal to implement the abovementioned learning design for a physics educator teaching a classroom for a semester with six hours per week. Instead of only lecturing for six hours, the classroom is transformed into a research hub where he/ she acts as a technical director rotating between the 12 groups with three students each. All groups are responsible for a specific physics domain, for example, Groups 11 and 12 focus exclusively on Experiment 4. This specialization allows students to develop cognitive depth in one area, which they will later simplify and explain to their peers during the Collaborative Presentation phase.

**Table 3**. Proposed Timeline for Using Gudeg in the Collaborative Project-based Learning

| Phase | Duration | Weekly Activity | Teacher & Student Focus |
|---|---|---|---|
| Phase 1: Initiation | Weeks 1–3 | 4h: Introduction and content review. | Teacher introduces the "Physics of Gudeg" framework. Students actively discuss with practitioners. |
| | | 2h: Engage discussion with food practitioners. | |
| Phase 2: Exploration | Weeks 4–10 | 2h: Theory/ Problem solving. | Groups work in parallel. Each group becomes an expert in their specific experiments. |
| | | 4h: Lab Days with Weekly Rotation. Week 1 ➔ Groups A-C do Exp 1, D-F do Exp 2, G-I do Exp 3, etc. | |
| Phase 3: Analysis | Weeks 11–14 | 3h: Data modeling | Students act as consultant learners, calculating errors and plotting graphs. |
| | | 3h: Writing the laboratory report. | |
| Phase 4: Presentation | Weeks 15–17 | 6h: Collaborative seminar. | Groups present findings to the Class and Gudeg practitioners. |
| Phase 5: Evaluation | Weeks 18 | 6h: Reflection & final assessment. | Assessing project findings with the national curriculum standards. |

**Conclusion**

This paper demonstrates that the traditional process of making Gudeg can be meaningfully integrated into physics education through a collaborative project-based teaching model. By mapping each stage of culinary production to core physics concepts, students explore the physics world through the lens of indigenous knowledge. The collaborative instruction is structured through five distinct phases: initiation and contextualization, exploration and experimentation, analysis and conceptualization, collaborative presentation, and reflection and evaluation. This structured approach ensures that the five experimental ideas presented using the Gudeg, ranging from the mass density analysis to the heat transfer concept, are not isolated entities, but rather a cohesive investigation into how physical laws govern cultural heritage.

The experimental implications of this paper provide a unique bridge between physics world and local knowledge. By analyzing the morphological study of the jackfruit's core versus its perianth specifically demonstrates how structural mechanics (specific mass) dictate cooking efficiency, providing Gudeg practitioners with scientific validation for their traditional techniques. The elasticity limits of the jackfruit scaffold through the Young's modulus analysis, students gain a contextual understanding of soft matter physics that goes beyond the textbook. Furthermore, experiments regarding conduction and convection in heat transfer also allow students to see the direct relationship between thermodynamics and food preservation.

As a final remark, the implications of this paper extend beyond the study of Gudeg, providing a replicable framework for embedding various forms of indigenous knowledge into physics education. By contextualizing the curriculum based on local practices, educators can make physics more accessible and relevant to students' daily lives while fostering a generation of consultant learners who value cultural ownership. Future research should focus on obtaining empirical data to measure the effectiveness of this model on intended learning outcomes and long-term retention. Expanding this framework to other cultural contexts or scientific subjects will further refine how global academic standards can harmoniously coexist with the preservation of local wisdom.

**Data availability statement**

The data that support the findings of this study are available upon reasonable request from the authors.